\documentclass[sn-mathphys,Numbered]{sn-jnl}% Math and Physical Sciences Reference Style
\usepackage{graphicx}%
\usepackage{multirow}%
\usepackage{amsmath,amssymb,amsfonts}%
\usepackage{amsthm}%
\usepackage{mathrsfs}%
\usepackage[title]{appendix}%
\usepackage{xcolor}%
\usepackage{textcomp}%
\usepackage{manyfoot}%
\usepackage{booktabs}%
\usepackage{algorithm}%
\usepackage{algorithmicx}%
\usepackage{algpseudocode}%
\usepackage{listings}%
\usepackage{csquotes}
\usepackage{hyperref}
\theoremstyle{thmstyleone}%
\theoremstyle{thmstyletwo}%

\theoremstyle{thmstylethree}%

\begin{document}

%%%%%%%%% INCLUDE NEW COMMANDS HERE
\newcommand{\actaa}{Acta Astron.}   % Acta Astronomica
\newcommand{\araa}{Annu. Rev. Astron. Astrophys.}   % Annual Review of Astron and Astrophys
\newcommand{\areps}{Annu. Rev. Earth Planet. Sci.} % Annual Review of Earth and Planetary Science
\newcommand{\aar}{Astron. Astrophys. Rev.} % Astrononmy and Astrophysics Review
\newcommand{\ab}{Astrobiology}    % Astrobiology
\newcommand{\aj}{Astron. J.}   % Astronomical Journal
\newcommand{\ac}{Astron. Comput.} % Astronomy and Computing
\newcommand{\apart}{Astropart. Phys.} % Astroparticle Physics
\newcommand{\apj}{Astrophys. J.}   % Astrophysical Journal
\newcommand{\apjl}{Astrophys. J. Lett.}   % Astrophysical Journal, Letters
\newcommand{\apjs}{Astrophys. J. Suppl. Ser.}   % Astrophysical Journal, Supplement
\newcommand{\ao}{Appl. Opt.}   % Applied Optics
\newcommand{\apss}{Astrophys. Space Sci.}   % Astrophysics and Space Science
\newcommand{\aap}{Astron. Astrophys.}   % Astronomy and Astrophysics
\newcommand{\aapr}{Astron. Astrophys. Rev.}   % Astronomy and Astrophysics Reviews
\newcommand{\aaps}{Astron. Astrophys. Suppl.}   % Astronomy and Astrophysics, Supplement
\newcommand{\baas}{Bull. Am. Astron. Soc.}   % Bulletin of the AAS
\newcommand{\caa}{Chin. Astron. Astrophys.}   % Chinese Astronomy and Astrophysics
\newcommand{\cjaa}{Chin. J. Astron. Astrophys.}   % Chinese Journal of Astronomy and Astrophysics (now RAA)
\newcommand{\cqg}{Class. Quantum Gravity}    % Classical and Quantum Gravity
\newcommand{\epsl}{Earth Planet. Sci. Lett.}    % Earth and Planetary Science Letters
\newcommand{\frass}{Front. Astron. Space Sci.}    % Frontiers in Astronomy and Space Sciences
\newcommand{\gal}{Galaxies}    % Galaxies
\newcommand{\gca}{Geochim. Cosmochim. Acta}   % Geochimica Cosmochimica Acta
\newcommand{\grl}{Geophys. Res. Lett.}   % Geophysics Research Letters
\newcommand{\icarus}{Icarus}   % Icarus
\newcommand{\ija}{Int. J. Astrobiol.} % International Journal of Astrobiology
\newcommand{\jatis}{J. Astron. Telesc. Instrum. Syst.}  % Journal of Astronomical Telescopes, Instruments, and Systems
\newcommand{\jcap}{J. Cosmol. Astropart. Phys.}   % Journal of Cosmology and Astroparticle Physics
\newcommand{\jgr}{J. Geophys. Res.}   % Journal of Geophysics Research
\newcommand{\jgrp}{J. Geophys. Res.: Planets}    % Journal of Geophysics Research: Planets
\newcommand{\jqsrt}{J. Quant. Spectrosc. Radiat. Transf.} % Journal of Quantitiative Spectroscopy and Radiative Transfer
\newcommand{\lrca}{Living Rev. Comput. Astrophys.}    % Living Reviews in Computational Astrophysics
\newcommand{\lrr}{Living Rev. Relativ.}    % Living Reviews in Relativity
\newcommand{\lrsp}{Living Rev. Sol. Phys.}    % Living Reviews in Solar Physics
\newcommand{\memsai}{Mem. Soc. Astron. Italiana}   % Mem. Societa Astronomica Italiana
\newcommand{\maps}{Meteorit. Planet. Sci.} % Meteoritics and Planetary Science
\newcommand{\mnras}{Mon. Not. R. Astron. Soc.}   % Monthly Notices of the RAS
\newcommand{\nat}{Nature} % Nature
\newcommand{\nastro}{Nat. Astron.} % Nature Astronomy
\newcommand{\ncomms}{Nat. Commun.} % Nature Communications
\newcommand{\ngeo}{Nat. Geosci.} % Nature Geoscience
\newcommand{\nphys}{Nat. Phys.} % Nature Physics
\newcommand{\na}{New Astron.}   % New Astronomy
\newcommand{\nar}{New Astron. Rev.}   % New Astronomy Review
\newcommand{\physrep}{Phys. Rep.}   % Physics Reports
\newcommand{\pra}{Phys. Rev. A}   % Physical Review A: General Physics
\newcommand{\prb}{Phys. Rev. B}   % Physical Review B: Solid State
\newcommand{\prc}{Phys. Rev. C}   % Physical Review C
\newcommand{\prd}{Phys. Rev. D}   % Physical Review D
\newcommand{\pre}{Phys. Rev. E}   % Physical Review E
\newcommand{\prl}{Phys. Rev. Lett.}   % Physical Review Letters
\newcommand{\psj}{Planet. Sci. J.}   % Planetary Science Journal
\newcommand{\planss}{Planet. Space Sci.}   % Planetary Space Science
\newcommand{\pnas}{Proc. Natl Acad. Sci. USA}   % Proceedings of the US National Academy of Sciences
\newcommand{\procspie}{Proc. SPIE}   % Proceedings of the SPIE
\newcommand{\pasa}{Publ. Astron. Soc. Aust.}   % Publications of the Astron. Soc. of Australia
\newcommand{\pasj}{Publ. Astron. Soc. Jpn}   % Publications of the Astron. Soc. of Japan (note no full stop following Jpn)
\newcommand{\pasp}{Publ. Astron. Soc. Pac.}   % Publications of the Astron. Soc. of the Pacific
\newcommand{\raa}{Res. Astron. Astrophys.} % Research in Astronomy and Astrophysics (formerly CJAA)
\newcommand{\rmxaa}{Rev. Mexicana Astron. Astrofis.}   % Revista Mexicana de Astronomia y Astrofisica
\newcommand{\sci}{Science} % Science
\newcommand{\sciadv}{Sci. Adv.} % Science Advances
\newcommand{\solphys}{Sol. Phys.}   % Solar Physics
\newcommand{\sovast}{Soviet Astron.}   % Soviet Astronomy
\newcommand{\ssr}{Space Sci. Rev.}   % Space Science Reviews
\newcommand{\uni}{Universe} % Universe

\newcommand{\micron}{$\mu$m}
\newcommand{\oi}{[O\,{\scriptsize I}]}
\newcommand{\feii}{[Fe\,{\scriptsize II}]}
\newcommand{\sii}{[S\,{\scriptsize II}]}
\newcommand{\neii}{[Ne\,{\scriptsize II}]}
\newcommand{\ilaria}[1] {\textit{\textcolor{red}{[IP: #1]}}}
\newcommand{\suzan}[1] {\textit{\textcolor{blue}{[SE: #1]}}}
\newcommand{\tracy}[1] {\textit{\textcolor{green}{[TB: #1]}}}
\newcommand{\sylvie}[1] {\textit{\textcolor{purple}{[SC: #1]}}}
%%%%%%%%%

\title[Nested Winds from Young Stars]{JWST/NIRSpec Reveals the Nested Morphology of Disk Winds from Young Stars}

\author*[1]{\fnm{Ilaria} \sur{Pascucci}}\email{pascucci@arizona.edu}

\author[2]{\fnm{Tracy L.} \sur{Beck}}\email{tbeck@stsci.edu}

\author[3]{\fnm{Sylvie} \sur{Cabrit}}\email{sylvie.cabrit@obspm.fr}

\author[1]{\fnm{Naman S.} \sur{Bajaj}}\email{namanbajaj@arizona.edu}

\author[4]{\fnm{Suzan} \sur{Edwards}}\email{sedwards@smith.edu}

\author[5]{\fnm{Fabien} \sur{Louvet}}\email{fabien.louvet@univ-grenoble-alpes.fr}

\author[6]{\fnm{Joan} \sur{Najita}}\email{joan.najita@noirlab.edu}

\author[1,7]{\fnm{Bennett N.} \sur{Skinner}}\email{skinnb1@mcmaster.ca}

\author[8,9]{\fnm{Uma} \sur{Gorti}}\email{uma.gorti-1@nasa.gov}

\author[10]{\fnm{Colette} \sur{Salyk}}\email{cosalyk@vassar.edu}

\author[11]{\fnm{Sean D.} \sur{Brittain}}\email{sbritt@clemson.edu}

\author[12]{\fnm{Sebastiaan} \sur{Krijt}}\email{s.krijt@exeter.ac.uk}

\author[2]{\fnm{James} \sur{Muzerolle}}\email{muzerol@stsci.edu}

\author[8]{\fnm{Maxime} \sur{Ruaud}}\email{maxime.ruaud@nasa.gov}

\author[13]{\fnm{Kamber} \sur{Schwarz}}\email{schwarz@mpia.de}

\author[13,14]{\fnm{Dmitry} \sur{Semenov}}\email{semenov@mpia.de}

\author[5,15]{\fnm{Gaspard} \sur{Duchene}}\email{gaspard.duchene@univ-grenoble-alpes.fr}

\author[16]{\fnm{Marion} \sur{Villenave}}\email{marion.villenave@unimi.it}

%\author[17]{\fnm{Leonardo} \sur{Testi}}\email{leonardo.testi@unibo.it}

\affil*[1]{\orgdiv{Lunar and Planetary Laboratory}, \orgname{The University of Arizona}, \orgaddress{\street{1629 E. University Blvd.}, \city{Tucson}, \postcode{85721}, \state{Arizona}, \country{USA}}}

\affil[2]{\orgdiv{Instruments Division}, \orgname{Space Telescope Science Institute}, \orgaddress{\street{3700 San Martin Drive}, \city{Baltimore}, \postcode{21218}, \state{Maryland}, \country{USA}}}

\affil[3]{\orgdiv{Observatoire de Paris}, \orgname{LERMA, CNRS}, \orgaddress{\city{Paris}, \postcode{75014}, \country{France}}}

\affil[4]{\orgdiv{Five College Astronomy Department}, \orgname{Smith College}, \orgaddress{\city{Northampton}, \postcode{01063}, \state{MA}, \country{USA}}}

\affil[5]{\orgdiv{University of Grenoble Alpes}, \orgname{CNRS, IPAG}, \orgaddress{\city{Grenoble}, \postcode{38000}, \country{France}}}

\affil[6]{\orgdiv{NOIRLab}, \orgname{NSF}, \orgaddress{\street{950 N Cherry Avenue}, \city{Tucson}, \postcode{85719}, \state{Arizona}, \country{USA}}}

\affil[7]{\orgdiv{Department of Physics and Astronomy}, \orgname{McMaster University}, \orgaddress{\city{Hamilton}, \postcode{L8S 4M1}, \state{Ontario}, \country{Canada}}}

\affil[8]{\orgdiv{Ames Research Center}, \orgname{NASA}, \orgaddress{\city{Moffett Field}, \postcode{94035}, \state{CA}, \country{USA}}}

\affil[9]{\orgdiv{Carl Sagan Center}, \orgname{SETI Institute}, \orgaddress{\city{Mountain View}, \postcode{94035}, \state{CA}, \country{USA}}}

\affil[10]{\orgdiv{Department of Physics and Astronomy}, \orgname{Vassar College}, \orgaddress{\street{124 Raymond Ave}, \city{Poughkeepsie}, \postcode{12604}, \state{NY}, \country{USA}}}

\affil[11]{\orgdiv{Department of Physics and Astronomy}, \orgname{Clemson University}, \orgaddress{\city{Clemson}, \postcode{29634}, \state{SC}, \country{USA}}}

\affil[12]{\orgdiv{Department of Physics and Astronomy}, \orgname{University of Exeter}, \orgaddress{\street{Stocker Road}, \city{Exeter}, \postcode{EX4 4QL}, \country{UK}}}

\affil[13]{\orgname{Max-Planck-Institut f\"{u}r Astronomie}, \orgaddress{\street{K\"{o}nigstuhl 17}, \city{Heidelberg}, \postcode{D-69117}, \country{Germany}}}

\affil[14]{\orgdiv{Department of Chemistry}, \orgname{Ludwig-Maximilians-Universit\"{a}t}, \orgaddress{\street{Butenandtstr. 5-13}, \city{M\"{u}nchen}, \postcode{D-81377}, \state{NY}, \country{Germany}}}

\affil[15]{\orgdiv{Department of Astronomy}, \orgname{University of California}, \orgaddress{\city{Berkeley}, \postcode{94720}, \state{CA}, \country{USA}}}

\affil[16]{\orgdiv{Dipartimento di Fisica}, \orgname{Universit\`a degli Studi di Milano}, \orgaddress{\street{Via Giovanni Celoria 16}, \city{Milano}, \postcode{20133}, \country{Italy}}}

%\affil[17]{\orgdiv{Dipartimento di Fisica}, \orgname{Università di Bologna}, \orgaddress{\street{Via Zamboni 33}, \city{Bologna}, \postcode{40126}, \country{Italy}}}

%#### AFFILIATIONS
%Sean D. Brittain
%orcid:0000-0001-5638-1330

% FORMATTING: https://www.nature.com/natastron/content
%Main text – up to 3,000 words, excluding abstract, Methods, references and figure legends.
%Abstract – up to 150 words, unreferenced.
%Display items – up to 6 items (figures and/or tables).
%References up to 50
%Article should be divided as follows:
%Introduction (without heading)
%Results
%Discussion
%Online Methods.

%%==================================%%
%% sample for unstructured abstract %%
%%==================================%%

% The current version of the abstract has 138 words
% The current version of the main text has 1,700 words
% Display figures we have 5, we can have up to 6
\abstract{
Radially extended disk winds could be the key to unlocking how protoplanetary disks accrete and how planets form and migrate. A distinctive characteristic is their nested morphology of velocity and chemistry. Here we report JWST/NIRSpec spectro-imaging of four young stars with edge-on disks in the Taurus star-forming region
that demonstrate the ubiquity of this structure. In each source, a fast collimated  jet traced by \feii\ is nested inside a hollow cavity within wider lower-velocity H$_2$ and, in one case, also CO ro-vibrational (v=1-0) emission. Furthermore, in one of our sources, ALMA CO\,(2-1) emission, paired with our NIRSpec images, reveals the nested wind structure extends further outward.
This nested wind morphology strongly supports theoretical predictions for wind-driven accretion and underscores the need for theoretical work to assess the role of winds in the formation and evolution of planetary systems. {\it In accordance with the Springer Nature publishing agreement, we can only post the submitted version of our article, prior to peer review. The version of record is published here: https://doi.org/10.1038/s41550-024-02385-7}
}


%%\pacs[JEL Classification]{D8, H51}

%%\pacs[MSC Classification]{35A01, 65L10, 65L12, 65L20, 65L70}

\maketitle

\section{Introduction}\label{sect:intro}
The assembly of stars and their planetary systems proceeds through an accretion disk where magneto-rotational instability (MRI, Balbus \& Hawley \cite{BH1991ApJ...376..214B})  was long favored to transport angular momentum outward, aiding inward accretion. Recent simulations, incorporating disk microphysics, challenge this view, showing MRI is suppressed in most of the planet-forming region ($\sim 1-20$\,au). Instead, radially extended magnetohydrodynamic (MHD) disk winds -- outflowing gas from a few scale heights above the disk midplane launched by magnetic and thermal pressures -- emerge as a viable mechanism for angular momentum removal and for enabling accretion \cite{Bethune2017A&A...600A..75B,Bai2017ApJ...845...75B,Gressel2020ApJ...896..126G}. This new wind-driven accretion scenario greatly impacts the evolution of the disk surface density \cite{Suzuki2016A&A...596A..74S}, affecting the inward drift of solids \cite{Taki2021ApJ...909...75T} and where planets may form and migrate to \cite{Ogihara2018A&A...615A..63O,Kimmig2020A&A...633A...4K}. Consequently, observations that can confirm the presence of these winds are crucial.

A unique feature of radially extended MHD winds is that flows are launched over a broad range of disk radii, hence outflow velocities, from near the gas co-rotation radius ($\sim 0.1$\,au) across much of the planet-forming region \cite{Wang2019ApJ...874...90W}. This contrasts with X-winds,
where stellar field lines couple to the disk at the co-rotation radius and launch fast ($\sim 150$\,km/s) MHD winds that spread widely \cite{Shu1994ApJ...429..781S}, and photoevaporative winds, which only develop where disk gas has enough thermal energy to escape the stellar gravitational field, beyond $\sim 1$\,au and with speeds $\leq 10$\,km/s \cite{Hollenbach1994ApJ...428..654H,Adams2004ApJ...611..360A}. Another distinctive characteristic of MHD winds is their nested morphology of velocity and chemistry throughout all stages of star-disk evolution: a fast $\sim 100$\,km/s jet, formed by the recollimation of the inner wind, lies inside atomic and molecular gas moving at sequentially lower velocities \cite{Cabrit1999A&A...343L..61C,Panoglou2012A&A...538A...2P}. X-winds can also collimate a jet \cite{Shu1995ApJ...455L.155S}. However, lower-velocity flows arise solely from  material entrained by the fast wind, which sweeps up gas near the disk surface  \cite{Matsuyama2009ApJ...700...10M} and, when an infalling envelope is present, also gathers envelope material near the disk's outer edge \cite{Cunningham2005ApJ...631.1010C}.
Conversely, in the photoevaporative scenario, the absence of recollimation means fast jets are not produced, and the wind's opening angle remains close to that at launch \cite{WG2017ApJ...847...11W}.

Current kinematic data on young ($\sim 0.1-10$\,Myr) stars with a mass greater than that of the infalling envelope -- Class~I and II sources \cite{Evans2009ApJS..181..321E} --  hint at the presence of radially extended MHD winds. Spatially unresolved spectra of optical forbidden emission like \oi{} and near-infrared emission of H$_2$ can show, in addition to a high velocity component ($\sim 50-300$\,km/s) tracing the base of a jet, a second lower velocity component with broad line widths interpreted as originating over disk radii $\sim 0.5-5$\,au \cite{Davis2001MNRAS.326..524D,Simon2016ApJ...831..169S,McGinnis2018A&A...620A..87M,Gangi2020A&A...643A..32G}.  A handful of spatially resolved outflows point to velocities which decrease outwards, with the narrow fast jet at the center of a wider-angle lower velocity flow (\cite{Pascucci2023ASPC..534..567P} and references therein).
However, the limited number of spatially resolved flows often coupled with modest sensitivity \cite{BB2019ApJ...884..159B,deValon2020A&A...634L..12D,Arulanantham2024arXiv240212256A} has thus far hindered a detailed mapping of the wind morphology across chemical species, leaving the ubiquity of the predicted nested structure by radially extended MHD winds uncertain.

\begin{table}[h!]
\centering
\caption{Literature source properties relevant to this study.}  \label{tab:sprop}
{\fontsize{7pt}{10pt}\selectfont
\begin{tabular}{lccccccccc}
\hline
 Source & RA(J2000) & Dec(J2000) & $M_{*}$ & $F_{\rm mm}$ & PA$_{\rm mm}$ & $i_{\rm mm}$ & $R_{\rm mm}$ & $R_{\rm ir}$ & Ref.   \\
 &  (h m s) & ($^\circ$ $'$ $''$) & (M$_{\odot}$) & (mJy) & ($^\circ$) & ($^\circ$)  & (au) & (au) & \\
\hline
FS~Tau~B &  04 22 00.7 & +26 57 32.5& 0.7 & 341 & 145 & 74 & 144 & 161 & 1,2\\
% FS~TauB = Haro 6-5B
HH~30 & 04 31 37.5 & +18 12 24.5& 0.5 & 55 & 121 & $>85$ & 130 & 217 &2,3 \\
IRAS~04302 & 04 33 16.5 & +22 53 20.4& 1.3-1.7& 268 & 175 & $> 84$ & 220 & &1,2\\
Tau~042021 & 04 20 21.4 & +28 13 49.2& 0.4& 124 & -16 & $> 85$ & 287 & 350 & 2,4\\
%TAU~042021 = 2MASS J04202144+2813491
\hline
\end{tabular}
}
\begin{tablenotes}
\item  {\bf Notes.} $M_{*}$ is inferred from the disk gas's Keplerian rotation, observed via well-resolved mm-wavelength spectro-imaging of molecular lines. $F_{\rm mm}$, PA$_{\rm mm}$, $i_{\rm mm}$, $R_{\rm mm}$ are the flux density, position angle, inclination, and disk radius from spatially resolved ALMA Band7 (0.89\,mm) continuum images, which trace millimeter-size grains. $R_{\rm ir}$ is the disk radius from optical/near-infrared scattered light images, thus probing small sub-micron grains well coupled with the gas.
\item {\bf References.} (1) \cite{Simon2019ApJ...884...42S}; (2) \cite{Villenave2020A&A...642A.164V}; (3) \cite{Louvet2018A&A...618A.120L}; (4) \cite{Duchene2023arXiv230907040D}
\end{tablenotes}
\end{table}

\section{Results}\label{sect:results}
We used the near-infrared spectrograph (NIRSpec) on the James Webb Space Telescope (JWST, \cite{Boker2022}) to acquire deep
spectro-imaging  of four edge-on disks with known jets \cite{Gomez1997AJ....114.1138G,EM1998AJ....115.1554E,Hartigan2007ApJ...660..426H,Villenave2020A&A...642A.164V,Duchene2023arXiv230907040D}. All sources belong to the nearby ($\sim 140$\,pc) and young ($\sim 1-2$\,Myr) Taurus star-forming region \cite{Luhman2018AJ....156..271L}, see Table~\ref{tab:sprop} for properties relevant to this study.
 While infrared spectral indices are often used to assign Classes to young stellar systems as proxies for their evolutionary stage \cite{Furlan2011ApJS..195....3F}, this method is unreliable for edge-on disks \cite{Crapsi2008A&A...486..245C}. Comparing the ALMA millimeter disk fluxes (Table~\ref{tab:sprop} and \cite{Villenave2020A&A...642A.164V}) with single-dish fluxes \cite{DiFrancesco2008ApJS..175..277D,Reipurth1993A&A...273..221R,AW2005ApJ...631.1134A,Andrews2013ApJ...771..129A} reveals that only IRAS~04302 is surrounded by a significant envelope. The other sources' fluxes are primarily from disk emission pointing to a more evolved evolutionary stage that aligns with the Class~II category.

\begin{figure}[!h]
\includegraphics[width=\textwidth]{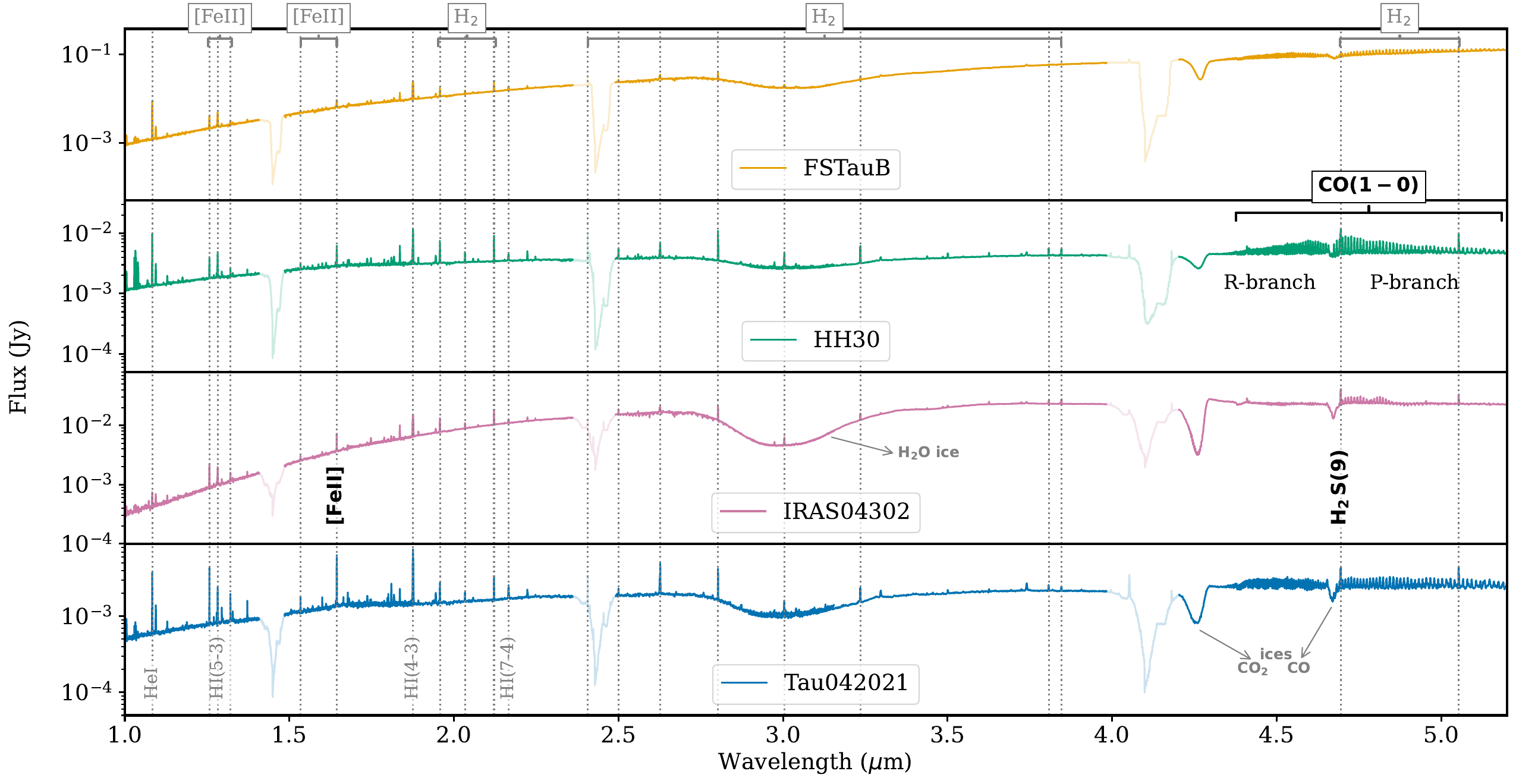}
\caption{Spectra integrated over the NIRSpec IFU. A few of the strongest lines are marked with gray dotted lines. All of these  lines are spatially extended. The transitions analyzed in this study are highlighted in boldface.}\label{fig:lines}
\end{figure}

NIRSpec IFU covers a field-of-view of $\sim 3 \times 3 ''$ with a pixel scale of $0.1''$ and, with our use of the three high-resolution gratings, delivers $0.95-5.27$\,\micron\ spectra at each pixel with a velocity sampling of $\sim 40$\,km/s
(see Sect.~\ref{sect:methods} for details on the observations and data reduction).
Multiple \feii{}, H, and H$_2$ lines are detected in the integrated spectra along with the P- and R-branches of the CO (v=1-0) ro-vibrational transition at $\sim 4.7$\,\micron\ (see Figure~\ref{fig:lines} highlighting some of the detected lines).  Here, we focus on the brightest \feii{} line at 1.644\,\micron , a known jet tracer \cite{Takami2006ApJ...641..357T}, the H$_2$ S(9) line at 4.695\,\micron , and the CO (v=1-0) band.  We selected the H$_2$ S(9) line, the brightest long-wavelength transition in our sample, due to significantly lower scattering from the disk surface relative to shorter wavelengths H$_2$ lines, like the often-studied 2.12\,\micron\ S(1) line. Its integrated flux is at most half that of the S(1) line (Table~\ref{tab:jetwindprop} and Sect.~\ref{sect:methods}), thereby combining the advantages of tracing the H$_2$ morphology closer to the disk with considerable relative brightness.
We note that our sources' H$_2$ 2.12\,\micron\   luminosities ($\sim 0.5-2 \times 10^{-6}$\,L$_\odot$) are representative for Class~II objects and about an order of magnitude lower than those of the brightest H$_2$  systems like DG~Tau~A   \cite{Gangi2020A&A...643A..32G}.
The similarity of the \feii{} fluxes in our sample (Table~\ref{tab:jetwindprop}) hints at similar jet mass loss rates. Considering the estimate for HH~30 of   $2 \times 10^{-9}$\,M$_\odot$/yr \cite{Bacciotti1999A&A...350..917B} and factoring in the typical jet/accretion ratio of approximately 0.1 \cite{Nisini2018A&A...609A..87N}, it is likely that our sources are accreting at rates exceeding $10^{-8}$\,M$_\odot$/yr. This inference is supported by the sole literature mass accretion rate available for our sample, the one from FS~Tau~B which is $\sim 1.5\times 10^{-7}$\,M$_\odot$/yr \cite{White2004ApJ...616..998W}.

\begin{figure}[!htb]%
\centering
\includegraphics[width=0.8\textwidth]{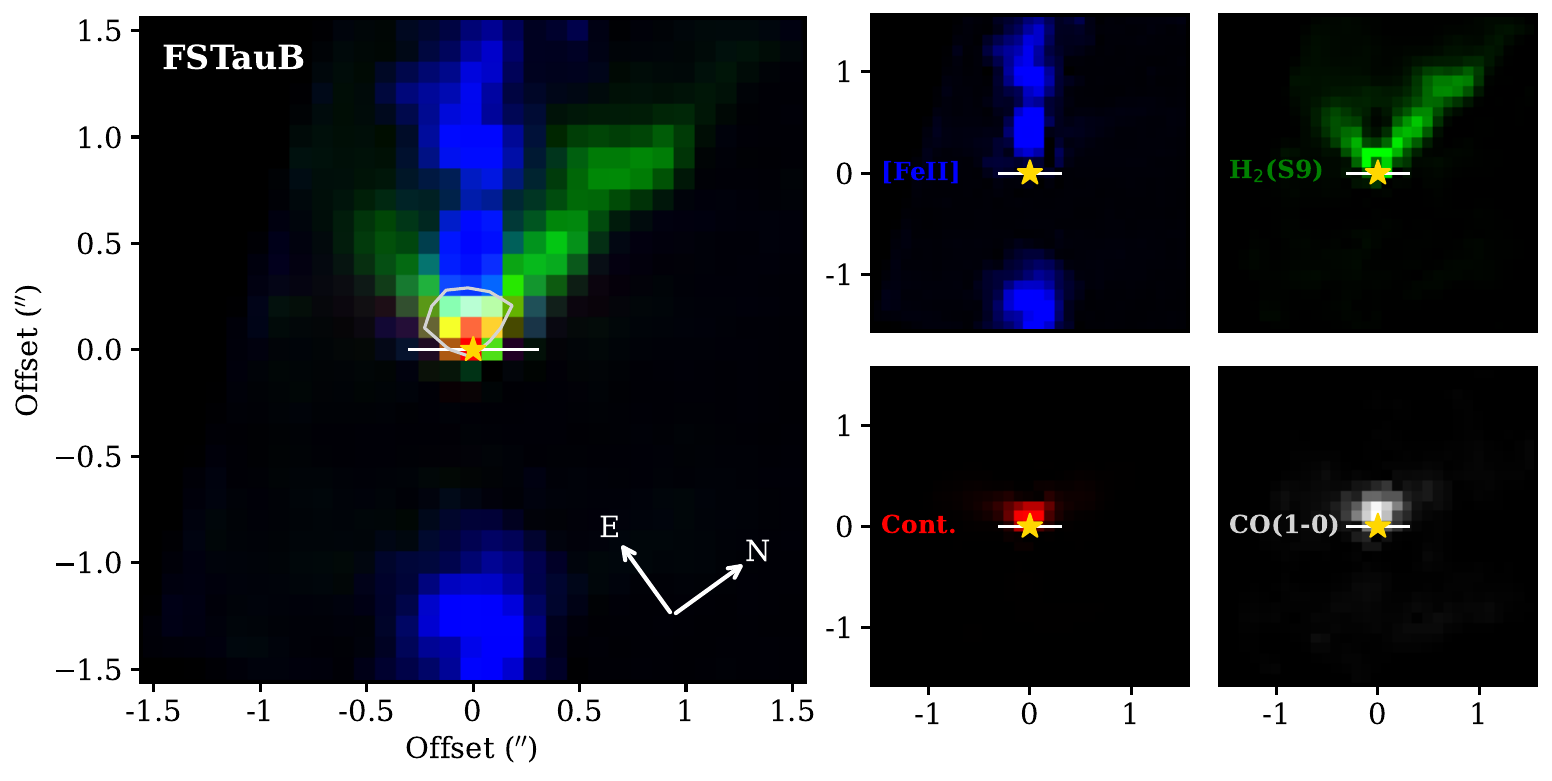}
\includegraphics[width=0.8\textwidth]{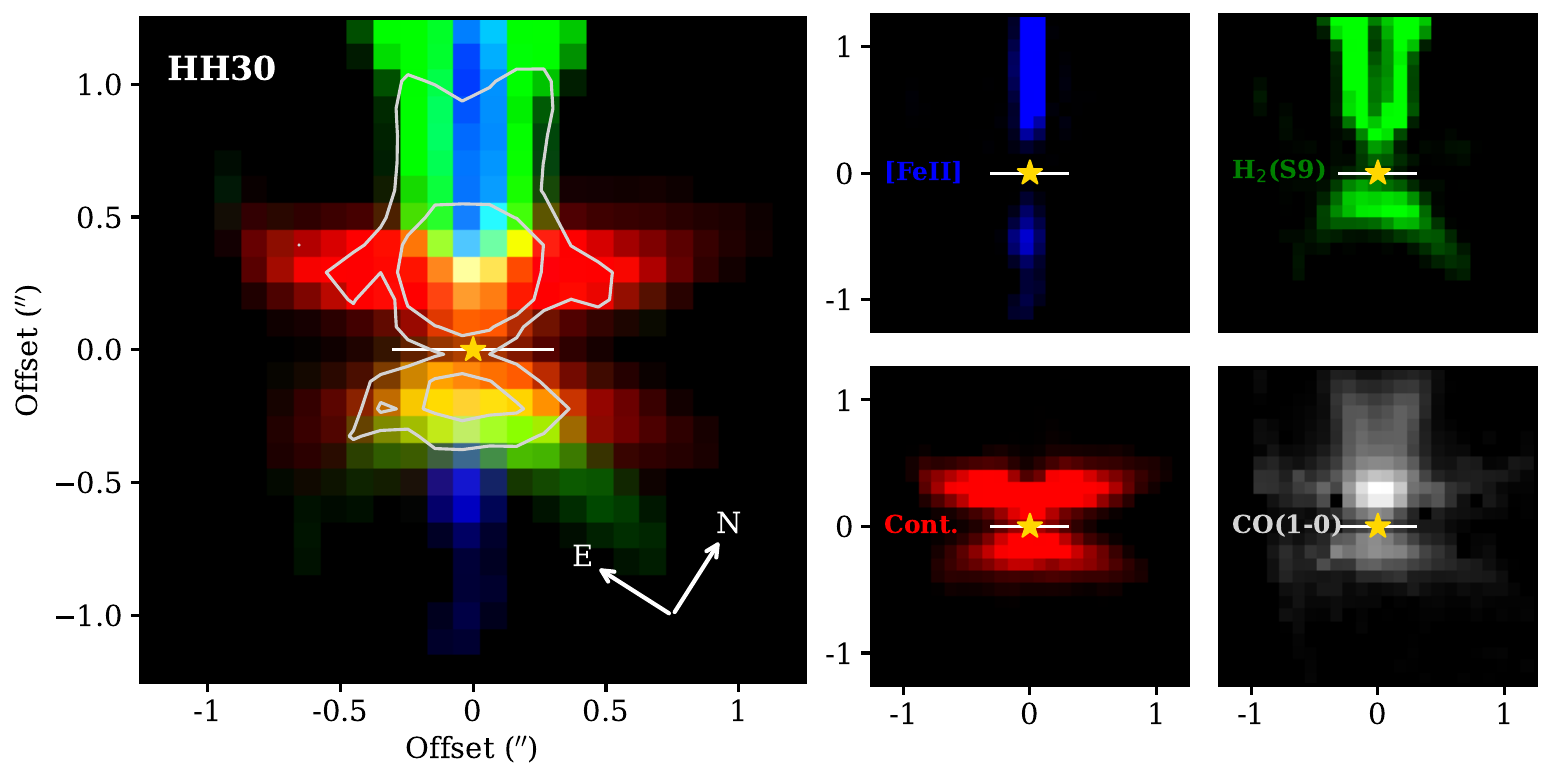}
\includegraphics[width=0.8\textwidth]{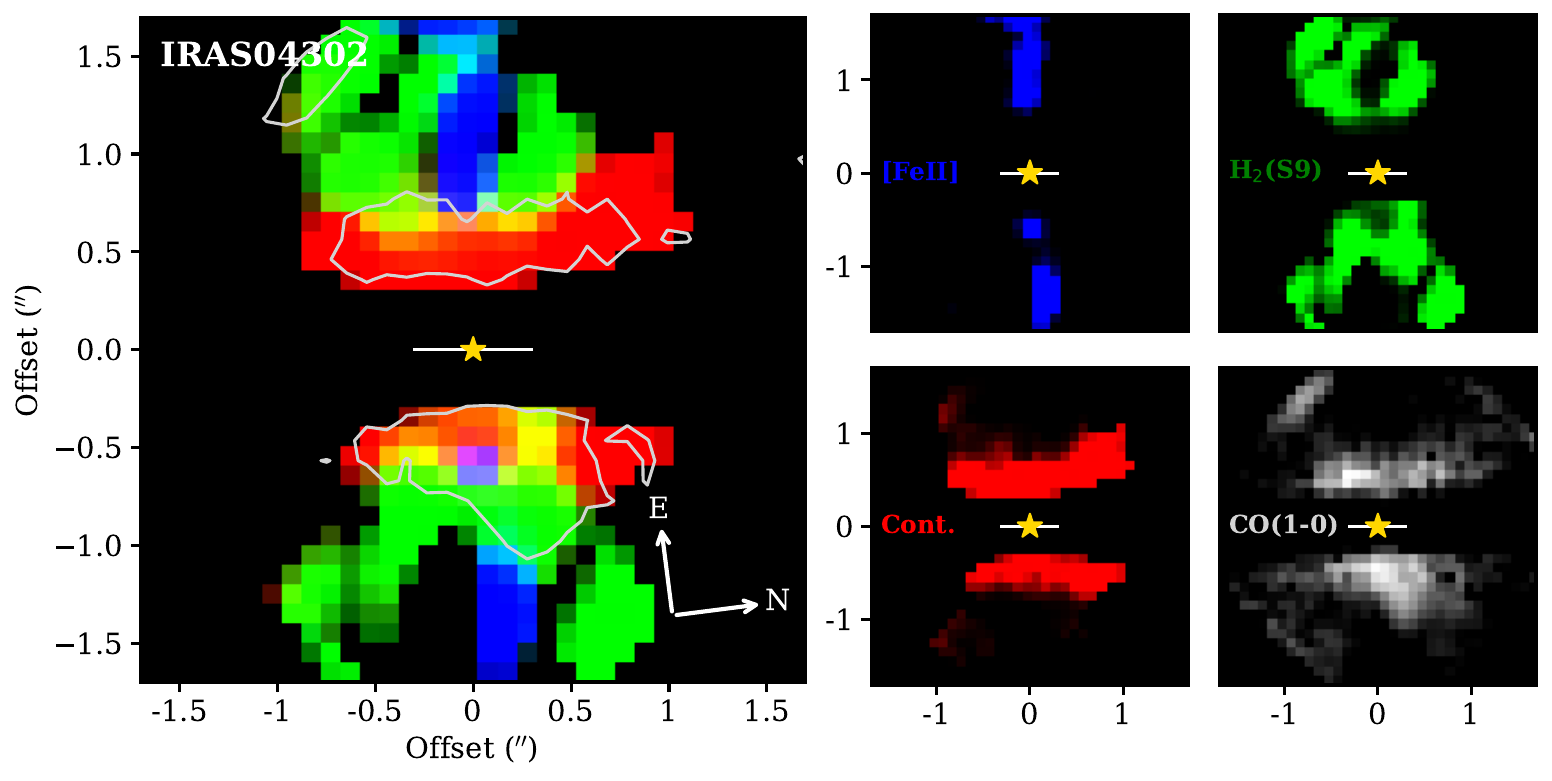}
\includegraphics[width=0.8\textwidth]{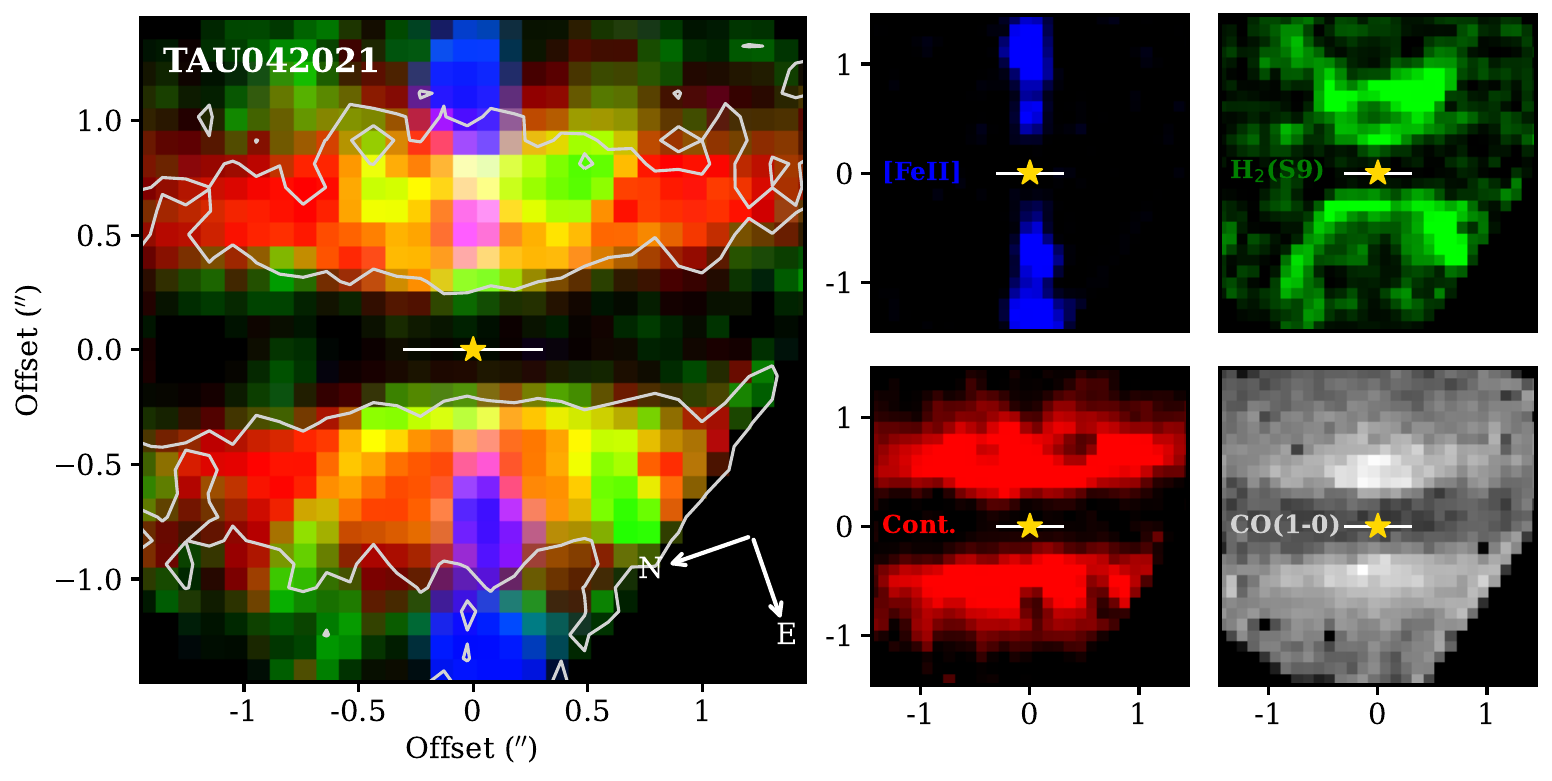}
\caption{Composite images NIRSpec: blue \feii{} emission at 1.644\,\micron\  – green H$_2$ emission at 4.695\,\micron\ - red continuum emission at the H$_2$ line. CO emission from $\sim 4.7-4.9$\,\micron\ is shown in the right bottom sub-panel in gray – the 50\% of the peak emission contour (and for HH~30 also the 30\%) is  shown in gray in the composite image. In every instance, the collimated jet traced in \feii\ is nestled inside the wider H$_2$ emission and, in the case of HH~30, also the CO (v=1-0) emission.}\label{fig:composite}
\end{figure}

The sub-panels of Figure~\ref{fig:composite} show PSF-deconvolved, continuum-subtracted line maps of the 1.644\,\micron\ \feii\ line, the H$_2$ S(9) line, along with the sum of the CO P-branch lines between $\sim 4.7-4.9$\,\micron\ (see Sect.~\ref{sect:methods} for details).  The continuum emission below the H$_2$ S(9) line, which traces scattered light by small grains at the disk surface,
is also displayed in one of the sub-panels. The main panels show three-color composite images combining the \feii\ and H$_2$ lines with the continuum emission.  Additionally, integrated CO P-branch contours are overlaid in gray. \feii{} traces the narrow jet emission and we have verified that its position angle is perpendicular to the ALMA millimeter disk (Tables~\ref{tab:sprop} and \ref{tab:jetwindprop}). We have also generated H$_2$ maps for the 2.12\,\micron\ and the 5.05\,\micron\ lines and found similar morphologies to the H$_2$ S(9) line which is thus representative for the molecular hydrogen gas.
In addition, we created velocity centroid maps for the \feii{} and H$_2$  lines and found that the NIRSpec resolution can discern small shifts between the blue and red lobe of the \feii\ jets (see Sect.~\ref{sect:methods}). Images in Figure~\ref{fig:composite} were  then rotated to align the blueshifted jet emission vertically at the top of each panel. The H$_2$ maps show no velocity structure, as expected if  molecular gas moves at lower speeds than the jet.

%%%%%%%%%%%%% RGB figures

In every source, the narrow \feii{} jet is nested within wider H$_2$ emission, which is always more vertically extended than the continuum emission at the same wavelength. While  the overall H$_2$ shape
has been seen in other sources \cite{BB2019ApJ...884..159B}, revealed for the first time is the ubiquity of a distinct central cavity, as expected if H$_2$ traces a disk wind emerging from a larger disk radius than the jet \cite{Panoglou2012A&A...538A...2P}.
%These data provide the first wind observations for FS~Tau~B and IRAS~04302.
Extended CO ro-vibrational emission is detected in all sources, yet uniquely in HH~30, this emission mirrors the H$_2$ conical morphology, albeit less vertically extended, with a central cavity surrounding the jet. This marks the first detection of a wind-like structure in the CO (v=1-0) band.

Significant brightness asymmetry relative to the disk plane is seen in the jet and wind features for two of the four sources: FS~Tau~B and HH~30. In both cases the more prominent emission is on the side of the disk where the jet is blueshifted. While this asymmetry has been known for the jets \cite{EM1998AJ....115.1554E,Hartigan2007ApJ...660..426H}, it is now seen in the winds as well. This discovery is particularly noteworthy for FS~Tau~B, marking the first such observation in its wind. For HH 30, our finding corroborates the asymmetry previously noted in the millimeter CO\,(2-1) wind emission \cite{Pety2006A&A...458..841P,Louvet2018A&A...618A.120L}.

\begin{figure}[h]%
\centering
\includegraphics[width=\textwidth]{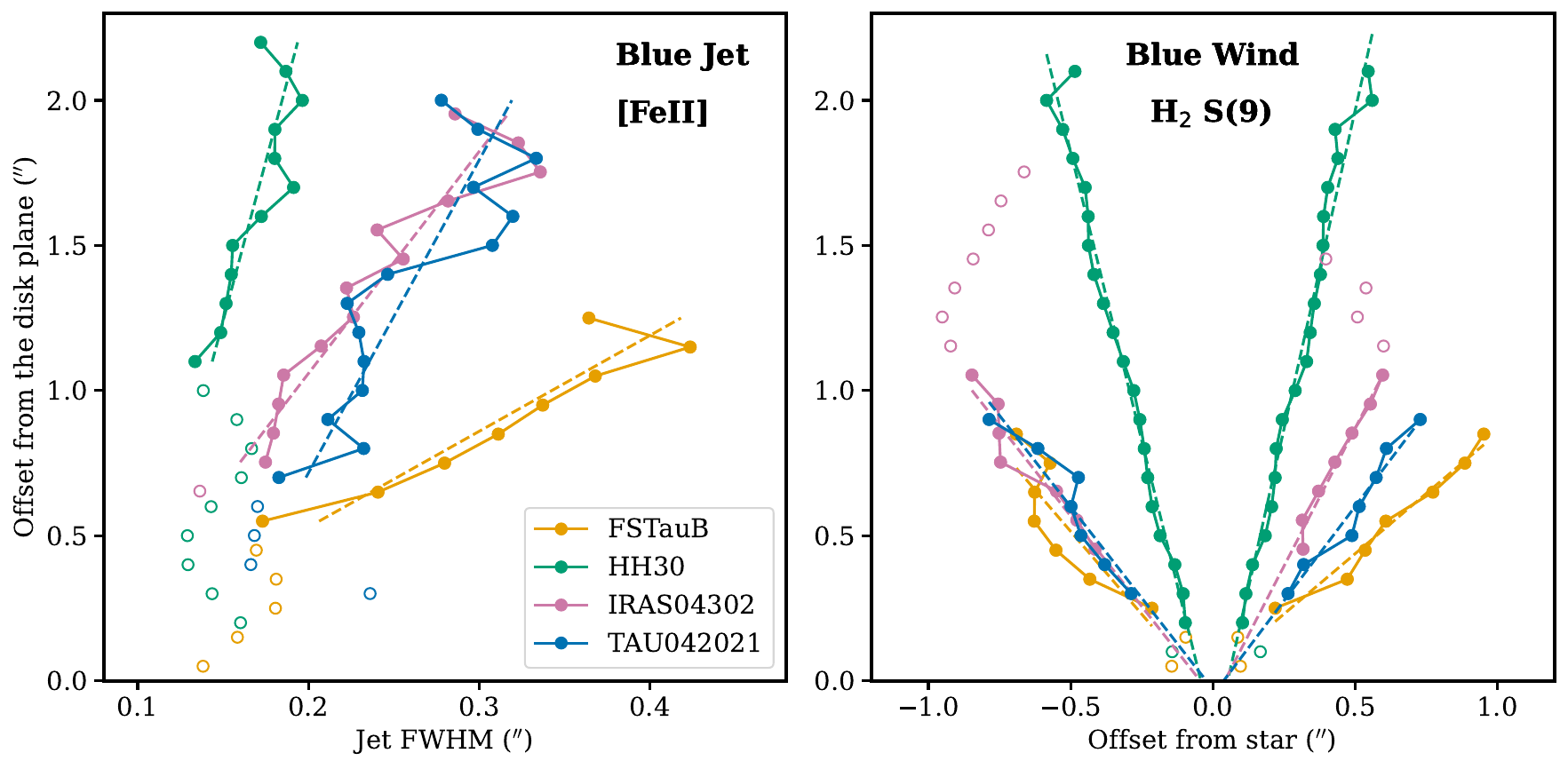}
\caption{Left panel: FWHM of the jet's blueshifted component as a function of distance from the disk plane. Right panel: Edges of the wind's blueshifted component vs. the disk's plane. In both panels filled symbols represent data points used in the linear fit to estimate the jet and the wind semi-opening angles and the geometric radii (see Table~\ref{tab:jetwindprop}). Best-fits are shown with dashed lines. Data points near the disk plane are not included in the fit due to inadequate sampling of the FWHM and scattering (see Sect.~\ref{sect:methods} for details). For IRAS~043202 we exclude points above 1.2$''$ as they clearly deviate from a wind-like morphology. HH~30 has the narrowest jet and wind in our sample. }\label{fig:jets_winds}
\end{figure}

For each source, we trace the \feii\ jet and H$_2$ wind emission to quantify their semi-opening angles ($\theta_{\rm j}$ and $\theta_{\rm w}$) and estimate the radii where the wind intersects the disk plane (geometric radii, $R_{\rm geo}$), see Sect.~\ref{sect:methods} for details and Table~\ref{tab:jetwindprop} for the results.  Jet semi-opening angles vary between $\sim 1.5^\circ-15^\circ$, consistently smaller than their corresponding wind semi-opening angles which range from $\sim 15^\circ$ to $50^{\circ}$. HH~30 stands out as having the most collimated jet and narrowest wind among our sources (Figure~\ref{fig:jets_winds}), although similarly narrow jets have been reported in other Class~II sources \cite{Maurri2014A&A...565A.110M,Erkal2021A&A...650A..46E}. The wind semi-opening angle we estimate for Tau~042021 is the same as that recently reported  in \cite{Arulanantham2024arXiv240212256A} using a lower spatial resolution H$_2$ S(2) map obtained with JWST/MIRI.
Our PSF-deconvolved images enable measuring the semi-opening angle of its jet, which was unresolved with MIRI. In addition, they allow placing stringent constraints on the intersection points of the H$_2$ cones with the disk plane. Specifically, we determine these intersections occur at $R_{\rm geo} \lesssim 15$\,au, significantly inside the disk radii.
If these cones trace MHD winds, the decreasing wind opening angle with height due to magnetic recollimation \cite{BP1982MNRAS.199..883B}, implies that each
$R_{\rm geo}$ defines an upper boundary for the actual wind-launching radius.

%==== Figure FWHM jet and winds' prop

%New Table removing dj because jets are known the be more extended than in the IFU field
\begin{table}[h!]
\centering
\caption{Inferred jet and wind properties. The \feii\ and H$_2$ lines we have selected are among the most prominent tracers of jets and winds, respectively.}  \label{tab:jetwindprop}
{\fontsize{7pt}{10pt}\selectfont
\begin{tabular}{lccc|ccc}
\hline
 Source & \multicolumn{3}{c|}{Jet: \feii\,1.644\,\micron} & \multicolumn{3}{c}{Wind: H$_2$\,S(9)\,4.695\,\micron} \\
       & Flux & PA & $\theta_{\rm j}$
       & Flux & $\theta_{\rm w}$ & $R_{\rm geo}$ \\
       & (erg/s/cm$^2$) & ($^\circ$) & ($^\circ$)
       & (erg/s/cm$^2$) & ($^\circ$)  & ($''$)   \\
\hline
FS~Tau~B & 4.1$\times 10^{-15}$  & 55$\pm$3 & 18.0$\pm$5.2 & 3.5$\times 10^{-15}$ & 49.9$\pm$13.8 & 0.07$\pm$0.11 \\
     & & & & &  50.1$\pm$5.0 & 0.01$\pm$0.06 \\
% FS~TauB = Haro 6-5B
HH~30 & 2.0$\times 10^{-15}$ & 32$\pm$2 & 1.4$\pm$0.9 & 1.5$\times 10^{-15}$ & 14.7$\pm$0.9 & -0.04$\pm$0.01 \\
      & & & & & 13.5$\pm$1.2 & 0.05$\pm$0.02 \\
IRAS~04302 & 3.0$\times 10^{-15}$ &87$\pm$2 & 9.0$\pm$1.9 & 3.9$\times 10^{-15}$ & 34.5$\pm$6.1 & -0.13$\pm$0.08 \\
      & & & & & 26.8$\pm$3.1 & 0.08$\pm$0.03  \\
%$^\dagger$      REDLOBE     &                &         & 0.3$\pm$1.6  &   &  35, 36 & -0.09, 0.11  \\
Tau~042021 & 3.3$\times 10^{-15}$ &72$\pm$2  & 4.9$\pm$1.7 & 4.5$\times 10^{-16}$ & 40.6$\pm$9.4 & -0.05$\pm$0.1\\
      & & & & & 42.3$\pm$4.9 & 0.03$\pm$0.06 \\
%$^\dagger$           &                &         & 9.1$\pm$2.7   &  & 41, 35 & -0.1, 0.4 \\
%TAU~042021 = 2MASS J04202144+2813491
\hline
\end{tabular}
}
\begin{tablenotes}
\item  {\bf Notes.} For each source we list the jet's semi-opening angle ($\theta_{\rm j}$) and the wind's semi-opening angle ($\theta_{\rm w}$) and geometric radius ($R_{\rm geo}$) for both lobes of the blue-shifted emission.
FS~Tau~B's disk is less inclined than the others, hence the disk plane and $R_{\rm geo}$ are less well determined (see Sect.~\ref{sect:methods}).
\end{tablenotes}
\end{table}

\section{Discussion and Conclusions}\label{sect:discussion}
Different types of disk winds, with distinct kinematic and morphological features, have been discussed in the literature (Sect.~\ref{sect:intro}). X-winds have been invoked to spin down the accreting star \cite{Shu1994ApJ...429..781S}, while photoevaporative winds might drive disk dispersal \cite{Alexander2014prpl.conf..475A}. Only radially extended MHD winds could solve the longstanding puzzle of how protoplanetary disks accrete (\cite{Lesur2023ASPC..534..465L,Pascucci2023ASPC..534..567P} for recent reviews).

The edge-on orientation of our systems provides a unique advantage to map out the structure of disk winds, with the bright flux of the central star effectively masked by the disk.
With NIRSpec IFU spectral imaging the morphological relation between atomic and molecular disk wind components on spatial scales of a few hundred 100\,au is revealed in striking detail in multiple systems for the first time. In all cases, a collimated jet traced by \feii{} is encased within the wider-angle emission from H$_2$, and in one case, also CO (v=1-0).  The near-infrared molecular emission extends well above 100\,au from the disk plane with a characteristic conical shape previously noted in a few other systems \cite{Schneider2013A&A...557A.110S,Agra-Amboage2014A&A...564A..11A,BB2019ApJ...884..159B,Arulanantham2024arXiv240212256A}. The  ubiquitous detection of a pronounced central cavity, marked by a notable absence of emission at the H$_2$ axial position, is a novel important finding.
Moreover, the molecular wind emission appears to be anchored at radial distances $\lesssim 15$\,au, well inside the disk radii. High-resolution ground-based spectroscopy of Class~I and II sources finds low-velocity H$_2$ 2.12\,\micron\ flows reaching speeds up to $\sim 20$\,km/s \cite{Davis2001MNRAS.326..524D,Gangi2020A&A...643A..32G}. Although H$_2$ velocities are not measured in our edge-on sources (Sect.~\ref{sect:results}), similarly low velocities likely characterize the molecular structures we are imaging.

\begin{figure}[h]
\centering
\includegraphics[width=0.9\textwidth]{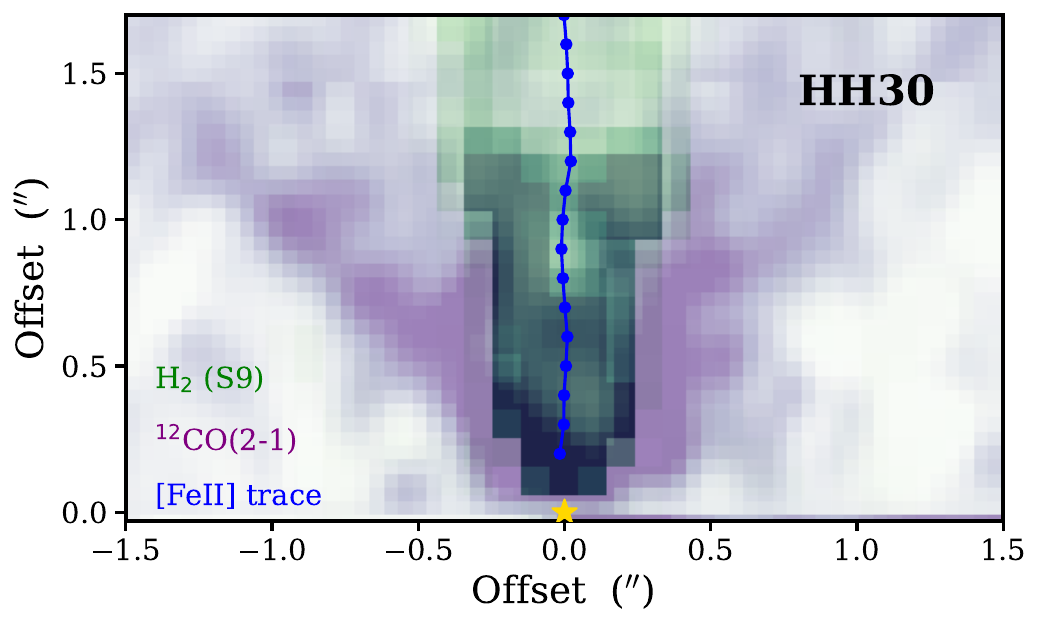}
\caption{Composite image of HH~30 highlighting the bueshifted wind components: NIRSpec H$_2$ (S9) emission in green (this work)  and ALMA CO\,(2-1) emission in purple (in the LSR velocity range $3-8$\,km/s and after removing the disk emission, \cite{Louvet2018A&A...618A.120L}). The trace of the \feii\ jet is also shown in blue.  With H$_2$ adjacent to and enclosed within the cooler CO wind traced by ALMA, this figure demonstrates the nested structure of disk winds.}\label{fig:NIRSpecALMA}
\end{figure}

In the case of HH~30, we can also compare the brightest component of the wind traced with NIRSpec in H$_2$ and CO with the one probed with ALMA in the CO\,(2-1) line, as shown in Figure~\ref{fig:NIRSpecALMA}. The CO millimeter emission traces gas flowing at $\sim 10$\,km/s and has a hollow cone shape, but with semi-opening angle $\sim 35^{\circ}$ \cite{Louvet2018A&A...618A.120L}, considerably wider than the $\sim$15$^{\circ}$ semi-opening angle (Table~\ref{tab:jetwindprop}) of the warmer H$_2$ and CO\,(v=1-0) cones. The latter nestle inside the millimeter cavity and, as expected, the average $R_{\rm geo}$ derived from H$_2$ ($\sim 6\pm2$\,au) falls below the 10\,au deduced from the CO\,(2-1) emission. Importantly, these distinctive hollow cone structures would likely be lost with less inclined disk orientations. If close to face on, arcs or rings might be seen \cite{Beck2008ApJ...676..472B,Pontoppidan2023arXiv231117020P}, but at most orientations the spatially extended molecular gas would simply appear diffuse, making it more difficult to pin down its connection to the disk.

%%%%%%%%%%%%% Composite HH30

Three of our systems appear to have dispersed most of their primordial envelopes, as indicated by the similarity of single-dish continuum millimeter fluxes and ALMA disk fluxes. Furthermore, for HH~30, recent ACA observations provide no evidence of a residual gas envelope (Louvet et al. submitted). The absence of an envelope disfavours the scenario where the observed wide-angle lower-velocity molecular flows would trace infalling envelope material swept-up and entrained by an X-wind \cite{Ai2024arXiv240202529A}. Additionally, our $R_{\rm geo}$ are more than an order of magnitude smaller than the disk radii which are the base of the X-wind swept-up shell \cite{Liang2020ApJ...900...15L}. The alternative of entrained lower-velocity material at the disk surface \cite{Matsuyama2009ApJ...700...10M} can be also excluded given the vertical extent of the H$_2$ emission detected with NIRSpec. Thus, while jet production by the X-wind cannot be ruled out, its role in generating the observed molecular flows seems very unlikely.

Photoevaporative winds produce molecular flows only beyond several au with velocities of just a few km/s and opening angles around 45$^\circ$ \cite{Komaki2021ApJ...910...51K}. Photoevaporative winds struggle to account for H$_2$ flows exceeding 5\,km/s and cannot match the 100\,au vertical extent seen in our NIRSpec data (see \cite{Rab2022A&A...668A.154R} for the predicted H$_2$ 2.12\,\micron\ emitting region). Moreover, the near-infrared H$_2$ and CO cones observed in HH~30 are too narrow for photoevaporative wind profiles (Figure~\ref{fig:windedges}) and the mass flux derived from CO ALMA data exceeds typical photoevaporative mass loss rates \cite{Louvet2018A&A...618A.120L}. Importantly, photoevaporative winds do not recollimate to produce fast axial jets, hence they would need to be coupled with an MHD wind to reproduce the narrow \feii{} emission seen in our sample. These inner winds might shield high-energy photons from the central star, diminishing their capacity to reach the outer disk and drive photoevaporative flows (Fig.~8 in \cite{Pascucci2023ASPC..534..567P}). The absence of \neii{} 12.8\,\micron\ low-velocity flows from stars with accretion rates $\gtrsim 10^{-8}$\,M$_\odot$/yr, akin to those of our sources (Sect.~\ref{sect:results}), coupled with low-velocity \oi\ tracing an inner wind, strongly suggests such shielding  \cite{Pascucci2020ApJ...903...78P}.

Among the proposed wind models, radially extended MHD winds are the only ones that can account for the ubiquity of the observed wind morphologies and nested structures. While the initial opening angles at the launch radii are $\ge 30^\circ$, the toroidal component of the magnetic field produces a gradual recollimation \cite{BP1982MNRAS.199..883B,Bai2017ApJ...845...75B,Gressel2020ApJ...896..126G} -- field lines close to the co-rotation radius contribute to a fast axial jet, while those further out can explain a range of flow opening angles depending on the scale probed by the observations. With temperature, ionization, and radiation field decreasing as the launch radius increases, a nested velocity structure and morphology are expected \cite{Cabrit1999A&A...343L..61C,Wang2019ApJ...874...90W}. H$_2$ and CO can survive in these winds and trace photoprocesses like those in  photodissociation regions and/or internal shocks \cite{Panoglou2012A&A...538A...2P}. Moreover, one-sided winds with respect to the disk plane, such as those seen in FS~Tau~B and HH~30, align with outcomes from recent global MHD simulations that account for the non-ideal Hall effect \cite{Bethune2017A&A...600A..75B,Bai2017ApJ...845...75B}.

The rich spectral information in this NIRSpec dataset, with numerous lines of \feii\ , H$_2$, and CO (Figure~\ref{fig:lines}), enables physical conditions in the wind components to be determined, including mass loss rates (Bajaj et al. in prep., Beck et al. in prep).  Together, the imaging and spectroscopy will clarify whether disk winds are indeed sufficient to drive disk accretion. The implications of wind-driven accretion extend far beyond clarifying how disks accrete. Wind-driven accretion profoundly impacts disk dynamics and planet formation:  it slows down the inward drift of solids, thus helping to overcome the radial drift barrier that hinders planet formation \cite{Taki2021ApJ...909...75T}; and it alters the migration of planetary embryos which enhances the survival of close-in super-Earths \cite{Ogihara2018A&A...615A..63O}. Additionally, MHD winds could transport high-temperature materials from the inner Solar System to the comet-forming region \cite{Giacalone2019ApJ...882...33G}. These effects could reshape our understanding of how planetary systems form and evolve.

\section{Methods}\label{sect:methods}

\noindent{\bf Observations and Data Reduction.}
Our sources were observed in September 2022 with the JWST NIRSpec IFU instrument as part of our Cycle~1 General Observer program  (PI~Pascucci, ID~1621). We adopted a 4-point dither pattern to improve the sampling of the PSF, one integration with 30 groups per dither, and integrated for a total on-source time of 30\,min per grating. We selected all three high-resolution gratings to cover the entire wavelength range from $\sim 0.98 - 5.2$\,\micron .

To reduce the data we utilized the JWST calibration pipeline \cite{Bushouse2023} version `1.11.2', which was made available on July 12, 2023. The initial stage, \texttt{Detector1Pipeline}, implements detector-level corrections on a group-by-group basis. We customized the \texttt{jump} step within this stage to detect `snowballs' in the data and flagged groups after jump with DN above 1000 and any groups within the first 50 seconds. Subsequently, the \texttt{Spec2Pipeline} was executed with default parameters: this step includes photometric calibration, flat field correction, and World Coordinate System assignment. Because of issues with the
default outlier detection and rejection step within the \texttt{Spec3Pipeline} stage, we ran a custom script to flag very large positive and negative pixels (J. Morrison, personal communication, 2023).  Following this, we ran the \texttt{Spec3Pipeline} with the modified `cal' files and created the image cubes with and without the `ifualign' mode, the latter avoids any further interpolation and is the preferred product for the analysis carried out in this paper. Finally, we inspected the cubes to verify that none of the scientific data were erroneously flagged. Additional information on the data reduction can be found in Bajaj et al. in prep. Finally, to properly compare NIRSpec IFU data across the full wavelength range in the dataset, the measured data cubes were deconvolved at each wavelength using a corresponding model PSF.  In doing this, the emission features sharpened to the same resolution level, with PSF broadening as a function of wavelength effectively removed.

The model PSF datacubes for the NIRSpec IFU mode were generated by de-convolving the corresponding commissioning observations of a point source (PID 1128) by the model NIRSpec data cube generated using the python based WebbPSF package \cite{Perrin2014SPIE.9143E..3XP} with the optical path differential (OPD) file that was appropriate for the date of the commissioning observation.  In doing this, we constructed a convolution kernel datacube that takes into account optical differences between the empirical data and PSF model calculations.  This kernel can hence be used to match any WebbPSF OPD calculation to an observed NIRSpec IFU dataset (Beck et al. in prep).  The PSF model used for this dataset study was constructed by using the WebbPSF OPD model for the time of the observations of HH 30 (26 Sep 2022; with NIRSpec wavefront error of 75nm rms), and convolving by our kernel to match the science data presented here.\\

\noindent{\bf Continuum-Subtracted Line Maps.}
To create line-only maps for the \feii{} and H$_2$ lines, we adopted the method outlined by \cite{Beck2008ApJ...676..472B}. First, we analyze the integrated spectrum from the entire NIRSpec IFU, fitting with {\tt scipy.optimize.curve\_fit} a Gaussian profile on top of the continuum to determine the total flux (reported in Table~\ref{tab:jetwindprop}) and the average line centroid and FWHM. Next, we extract the flux at each spaxel, fit a Gaussian+continuum, and evaluate if the line is detected by comparing the standard deviation of the data pre- and post-fit. Where a line is detected, we calculate the area of the Gaussian for the line-only map and store the centroid to determine which portion of the jet/wind is blueshifted.  Where no line is detected, we re-fit only the continuum and set a three-sigma upper limit for the line-only map using the standard deviation of the continuum-subtracted data and the average FWHM. The continuum-only map is derived by averaging the continuum across the line centroid$\pm$FWHM. Although not shown here, we have also generated line-only maps for the H$_2$ 2.12\,\micron\ and 5.05\,\micron\ transitions. Their integrated fluxes are $[4.9,3.7], [3.2,0.9], [4.1,1.9], [0.8,0.4]\times 10^{-15}$\,erg/s/cm$^2$ for FS~Tau~B, HH~30, IRAS~04302, and Tau~042021, respectively. Their morphologies are the same as that of the H2 S(9) line which we focus on here because it suffers less than the 2.12\,\micron\ from scattering close to the disk plane and it is brighter than the 5.05\,\micron\ line.

To construct the continuum-subtracted CO(v=1-0) map, we expand the method above by fitting, at each spaxel, multiple Gaussian profiles across various wavelength segments, all sharing a common continuum. We utilize the $^{12}$CO and $^{13}$CO line list from the HITRAN database \cite{Gordon2022JQSRT.27707949G}, supplemented by additional lines within the CO(1-0) fundamental band\footnote{\url{http://hebergement.u-psud.fr/edartois/jwst_line_list.html}}.
The line-only map is generated by summing the areas of all detected $^{12}$CO lines. In spaxels where lines are not detected, we provide a three-sigma upper limit, calculated using the standard deviation of the data after continuum subtraction and the average $^{12}$CO FWHM from fitting lines in the total NIRSpec IFU spectrum. \\

\noindent{\bf Jet and Wind Morphologies.}
To measure the jet position angle (PA), we use the continuum-subtracted \feii\ 1.644\,\micron\ map generated with the default rotation in the STScI pipeline script so that North is pointing up along the y axis.
For each pixel along the x-axis, we extract the corresponding y-array, smooth it, and find its peak value (y-peak). We then perform a linear fit of the x values and the identified y-peaks using the {\tt sklearn RANSACRegressor}. The jet PA is the angle between this best-fit line and the direction of North. We repeat the extraction, peak identification, and linear fit across different sets of (x,y) points  that exhibit jet emission. Next, we analyze the distribution of the jet PAs and compute the mode,  the most frequently occurring PA, and the standard deviation. These are the values reported in Table~\ref{tab:jetwindprop}. Note that in all cases the jet PA is perpendicular to the disk PA within the quoted uncertainties.

We determine the blueshited side of the jet from the Gaussian line centroid map which is constructed as discussed in the previous subsection. We then rotate all line-only and continuum maps to an angle that aligns the blueshifted component of the jet with the upper part of the y-axis on the panel. The jet's emission pinpoints the star's position along the x-axis. However, locating the star on the y-axis is more challenging due to the systems being observed close to edge-on. This orientation results in the circumstellar disk obscuring partially or completely the star's emission.
For HH~30, IRAS~04302, and Tau~042021, which have a disk inclined by more than $80^\circ$, the continuum map traces the flared disk surface, with no emission near the disk plane. For these cases,  we extract a profile along the jet emission and identify the midpoint of the non-emissive, or \enquote{dark}, continuum valley.  This midpoint is assigned as the star's y-axis location as well as the disk plane. In FS~Tau~B, with the disk inclined at $\sim 70^\circ$, the continuum emission appears point-like (see Figure~\ref{fig:composite}) and is likely to trace scattered light in proximity to the central star. We assign the star's y-axis position to to be the peak of the continuum emission.

We use the graduate broadening of the jet emission away from the disk plane to estimate the jet semi-opening angles ($\theta_{\rm j}$) reported in Table~\ref{tab:jetwindprop}. First, we perform horizontal cuts across the \feii{} rotated line-only images, fit a Gaussian profile, and measure its FWHM.  Next, we assess potential artificial broadening of the jet's FWHM due to scattering near the disk plane, along with any insufficient FWHM sampling. For FS~Tau~B, IRAS~04302, and Tau~042021, we noted that the peak of the jet emission is spread over two columns. Consequently, any FWHM measurement below 1.7 pixels (equivalent to 0.17$''$) is considered to be undersampled. This is evidenced by the near-constant jet FWHM for FS~Tau~B and Tau~042021 at distances below 0.5$''$ from the disk plane, as shown in Figure~\ref{fig:jets_winds}. Hence, we have excluded any data points with FWHM$< 0.17''$ when determining the jets' semi-opening angles. In contrast, the peak jet emission in HH~30 is more narrowly confined along a single column, which mitigates the sampling issue. For this source, we further compared our FWHM vs distance with that obtained from higher resolution STIS spectral images of the \sii\ 6731\,\AA\ line \cite{Hartigan2007ApJ...660..426H} and determined that the \feii\ jet becomes resolvable at distances greater than 1$''$ from the disk plane. Only these resolvable points were used in our analysis.
An Ordinary Least Squares fitting was then applied to the data points unaffected by scattering or poor sampling, which are represented by filled circles in Figure~\ref{fig:jets_winds}. The jet semi-opening angle ($\theta_{\rm j}$) is calculated from the slope of the best fit relation, this value along with its uncertainty are provided in Table~\ref{tab:jetwindprop}. For FS~Tau~B and HH~30, where jet semi-opening angles have previously been reported in the literature \cite{Mundt1991A&A...252..740M,Hartigan2007ApJ...660..426H}, our values align with the published ones, falling  within the estimated uncertainties.

\begin{figure}[h]%
\centering
\includegraphics[width=0.9\textwidth]{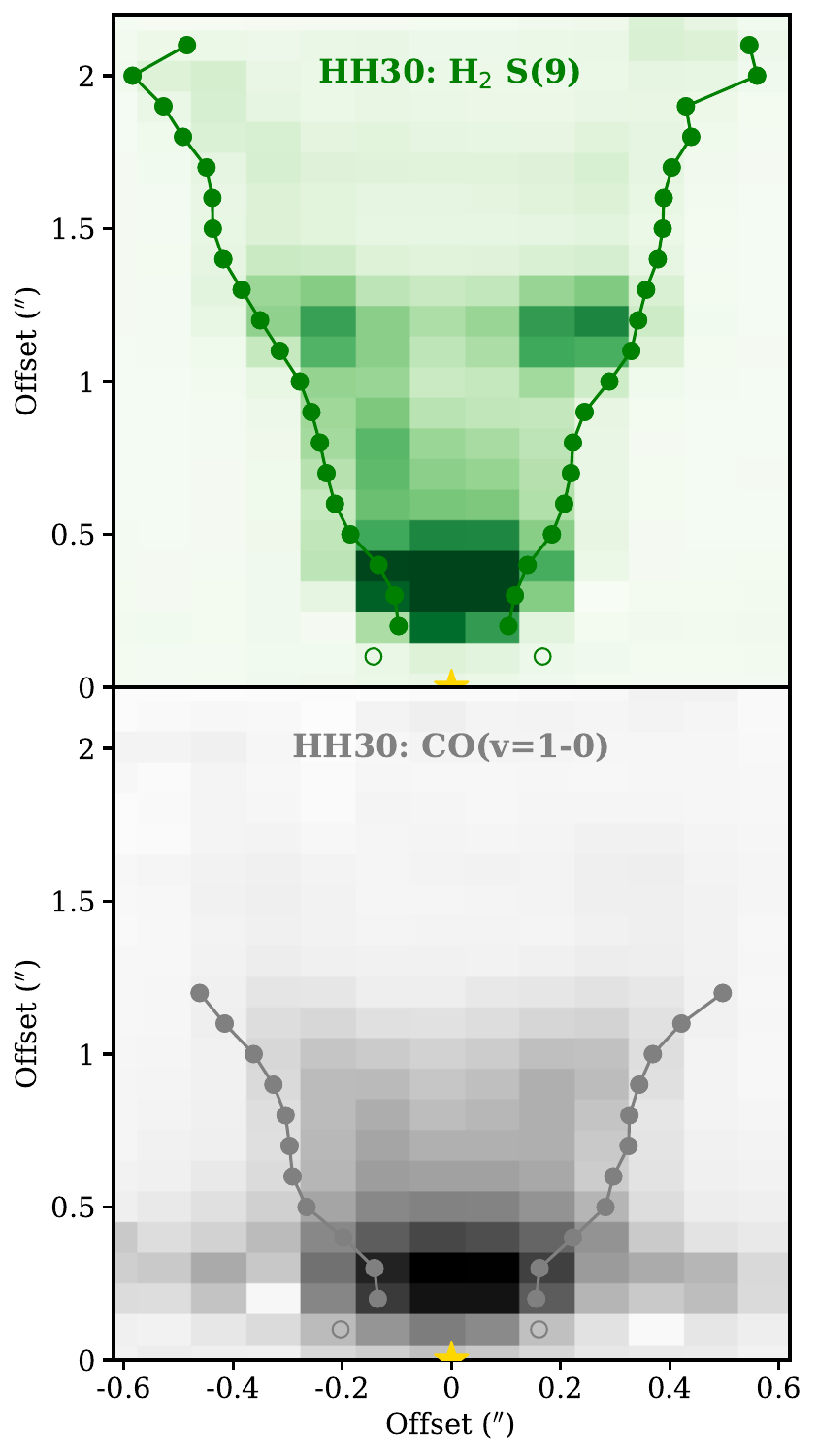}
\caption{Figure demonstrating our custom edge detection technique applied to HH~30.  Green empty circles are the identified edges of the H$_2$ S(9) emission shown here in greyscale. The trace of the \feii\ jet is also shown in blue.}\label{fig:windedges}
\end{figure}

The wind semi-opening angles ($\theta_{\rm w}$) and geometric radii ($R_{\rm geo}$) reported in Table~\ref{tab:jetwindprop} are measured on the rotated H$_2$ (S9)  line-only image. To trace the wind's outer contours, we developed
two approaches. The first one follows  \cite{Habel2021ApJ...911..153H} and relies on calculating the second-order derivative of horizontal cuts across the wind emission and finding its zero values $-$ these values indicate the inflection points that define the wind's edge. However, for complex patchy emission like that displayed by IRAS~04302 the method finds too many inflection points that need to be sorted out manually. As such, we developed a more robust aproach that relieas on fitting Gaussian profiles, typically two but see below, to the same horizontal cuts. We have verified on HH~30 that the wind's edges from the first method are equivalent to the Gaussian centroids plus or minus, depending on the wind's location with respect to the central axis, its FWHM.   For FS~TauB and HH30 one Gaussian$\pm$FWHM is used when approaching the disk plane. The Gaussian fitting method is applied to all sources and used to determine the wind's properties. Figure~\ref{fig:windedges} demonstrates this technique on HH~30.
To evaluate if the wind's shape exhibits symmetry around the central axis, we analyze the edges on either side separately. We fit the edges vs. the distance from the star using an Ordinary Least Squares fitting routine. The semi-opening angles ($\theta_{\rm w}$) of the wind are then calculated as the angles between the best-fit lines and the vertical y-axis. The geometric radii ($R_{\rm geo}$) are determined by the intersection points of these best-fit lines with the disk's plane, marked by the star's location on the y-axis. Uncertainties for $\theta_{\rm w}$ and $R_{\rm geo}$ are also calculated and reported in Table~\ref{tab:jetwindprop}.\\

\noindent{\bf Data Availability.}
The JWST data used in this paper can be found at the Mikulski archive for Space Telescope under programs ID~1621 and 1128. The first program covers the four edge-on disks while the second has commissioning data to generate the NIRSpec PSF. Both raw data and fully processed pipeline data can be downloaded from the archive. The spectra integrated over the NIRSpec IFU are available through figshare (DOI: 10.6084/m9.figshare.25396258). Upon request, the first author will provide any of the fits files used for this research. \\

\noindent{\bf Code Availability.}
The data were reduced with the JWST calibration pipeline version 1.11.2. Upon request, the first author will provide the python scripts to analyze the data and generate figures.

\backmatter

\bmhead{Acknowledgments}
This work is based on observations made with the NASA/ESA/CSA James Webb Space Telescope. The data were obtained from the Mikulski Archive for Space Telescopes at the Space Telescope Science Institute, which is operated by the Association of Universities for Research in Astronomy, Inc., under NASA contract NAS 5-03127 for JWST. These observations are associated with the GO Cycle 1 program 1621.
I.P. and N.B. acknowledge partial support from NASA/STScI GO grant JWST-GO-01621.001. D.~S. acknowledges support from the European Research Council under the Horizon 2020 Framework Program via the ERC Advanced Grant Origins 83 24 28 (PI: Th. Henning). G.D. acknowledges support from the European Research Council (ERC) under the European Union's Horizon Europe research and innovation program (grant agreement No. 101053020, project Dust2Planets, PI F.M\'enard).  M.V. acknowledges the support from the European Research Council (ERC) under the European Union’s Horizon Europe Research \& Innovation Programme under grant agreement no. 101039651 (DiscEvol). I.P. thanks D. Deng and F. Long for an initial exploration of the ALMA data for our sources. We acknowledge the use of the following packages: api, astropy, numpy, scipy, sklearn, matplotlib, and pandas.

%\begin{appendices}
%\section{Section title of first appendix}\label{secA1}
%An appendix contains supplementary information that is not an essential part of the text itself but which may be helpful in providing a more comprehensive understanding of the research problem or it is information that is too cumbersome to be included in the body of the paper.

%%=============================================%%
%% For submissions to Nature Portfolio Journals %%
%% please use the heading ``Extended Data''.   %%
%%=============================================%%

%%=============================================================%%
%% Sample for another appendix section			       %%
%%=============================================================%%

%% \section{Example of another appendix section}\label{secA2}%
%% Appendices may be used for helpful, supporting or essential material that would otherwise
%% clutter, break up or be distracting to the text. Appendices can consist of sections, figures,
%% tables and equations etc.

%\end{appendices}

%%===========================================================================================%%
%% If you are submitting to one of the Nature Portfolio journals, using the eJP submission   %%
%% system, please include the references within the manuscript file itself. You may do this  %%
%% by copying the reference list from your .bbl file, paste it into the main manuscript .tex %%
%% file, and delete the associated \verb+\bibliography+ commands.                            %%
%%===========================================================================================%%

\bibliography{sn-bibliography}% common bib file
%% if required, the content of .bbl file can be included here once bbl is generated
%%\input sn-article.bbl

\end{document}